# Delayed $^{160}$Tb radioactivity buildup due to $^{159}$Tb($n,^2n$) nuclear reaction products transformation and subsequent fusion

Ihor M. Kadenko [1, 2, a], Nadiia V. Sakhno [1], Oleksandr M. Gorbachenko [2], Anastasiia V. Synytsia [2]

[1] International Nuclear Safety Center of Ukraine, Taras Shevchenko National University of Kyiv, St. Volodymyrska, 64/13, Kyiv, 01601, Ukraine

[2] Department of Nuclear and High Energy Physics, Faculty of Physics, Taras Shevchenko National University of Kyiv, St. Volodymyrska, 64/13, Kyiv, 01601, Ukraine

**Abstract:** This paper deals with the formation of a bound dineutron in the outgoing channel of the $^{159}$Tb($n,^2n$) $^{158g}$Tb nuclear reaction followed by assumed transformations of the reaction products $^{158g}$Tb and $^2n$. Such nuclear processes were studied in detail from the point of view of $^{160}$Tb/$^{160}$Dy/$^{160}$Ho amount of nuclei versus time dependence. Some signs of fusion process between heavier nuclei ($^{158}$Tb and/or $^{158}$Gd) and the deuteron, that is a bound dineutron decay product, were detected as unexpected increasing of 879.38 keV gamma-ray peak count rate due to $^{160}$Dy gamma-transitions. The mathematical model, including three systems of differential equations, was developed to describe the experimental data. This development requires a reasonable estimate of the half-life of a bound dineutron, which was found to be equal 5,877 s as an upper limit. We mathematically modelled the experimentally observed delayed in time build-up of the $^{160}$Tb radioactivity with a maximum at about 495 d since the neutron irradiation completion of the Tb sample, based on the similarity with the parent – daughter nuclei radioactivity decay and nuclear accumulation processes.

## 1. Introduction

A bound dineutron occupies a separate place in nuclear science as one of the three representatives of possible two-nucleon systems including its charge symmetric partners (the deuteron and the diproton) and consisting of the two neutrons only. Bound dineutrons have been the subject of scientific hunt since the mid of last century [1, 2, 3]. Among these possible two-nucleon bound configurations according to the hierarchy of their masses, the deuteron is most likely to be the lightest nucleus, potentially allowing a corresponding beta decay of the dineutron or the diproton with the formation of the deuteron instead. However, for decades such two identical nucleon bound nuclei were considered as non-existing due to the Pauli Exclusion Principle. At the same time, some theoretical studies do not rule out at least a bound dineutron [4, 5, 6]. Experimental searches for a fully neutral two-neutron system such as a bound dineutron were targeted at two different atomic mass regions: light (A<30) and heavier nuclei (A>55). While some experiments were performed in order to detect the possible emission of a bound dineutron in fission reactions [7], others were devoted to the introduction of two neutrons consequentially, or via the dineutron state in the ($^3$H, $p$) nuclear reaction into Rh and Co nuclei [8]. Much more intensive research was dedicated to light exotic and neutron-rich nuclei to study either dineutron correlations or dineutron decay [9, 10, 11, 12, 13]. For light nuclei $^6$He, $^8$He, $^{11}$Li, $^{14}$Be, $^{17}$B based on experimental data on interaction cross sections, being larger than those of their neighbors; expected abnormally large radii; smaller separation energies of the last two neutrons and the density distribution of neutrons, reflecting a very long tail [14], it was suggested that a bound dineutron may exist on the nuclear core surface, constituting what was called a neutron "halo" or neutron "skin" comprising the volume of $^6$He, $^8$He, $^{11}$Li, $^{14}$Be, $^{17}$B and other nuclei. Also, a dineutron emission is claimed in the decay of $^{16}$Be [11] and $^{26}$O [13]. However, it is not obvious at all how one can deduce from these observations the existence of a bound dineutron [15, 16, 17]. In order to consider a possible existence of a bound dineutron for some light nuclei, the authors in [18] revived an idea due to Migdal [4], suggesting that while the dineutron itself is known to be unbound by 66 keV, being localized "… in the field of a nucleus it may act as a bound pair weakly bound to the nucleus. In other words, the dineutron may exist as a bound system on the nuclear surface…" However, it was a misinterpretation of an original Migdal's idea, according to which the dineutron may be localized within the potential well not on, but near the nuclear surface of a heavy nucleus. The last fact is extremely important in order to avoid solving the three-body problem and considering the heavy nucleus with 100 <A< 200 [19] as a source of the potential field. Possibly due to this feature no bound dineutrons were detected in any reactions on light nuclei so far. One more direction to search for a bound dineutron deals with a new technique of precision electron induced hard proton knockout from $^3$H [16], but up to now the authors could not identify a signature for a bound dineutron with the binding energy, close enough to the binding energy of the deuteron. Then a new experiment is needed to consider the lower binding energies of a bound dineutron [20, 21].

---

[a] Corresponding author.
E-mail address: imkadenko@univ.kiev.ua (I.M. Kadenko).

Observation of a new nuclear process with the formation of the dineutron in the output channel in the $^{159}$Tb(n, $^2$n)$^{158}$Tb nuclear reaction was declared in [19] for the first time and validated by the statistically and systematically significant detection of a bound dineutron in the same type of nuclear reaction, but with an $^{197}$Au nucleus [22] in the input channel. Both these works confirmed the existence of a new nuclear reaction type and channel [23], essentially different in their properties from the commonly known nuclear reaction mechanism, for which all the reaction products in the outgoing channel are well separated in space and leave each other in time. Migdal predicted the formation of the dineutron in the output channel of a nuclear reaction, when two neutrons combine into a bound system due to the existence of additional bound states within the potential well of a heavy nucleus, but outside its volume [4]. In line with this prediction, it is only possible to directly observe one of the two reaction products, unequivocally prescribing the existence of a bound dineutron, as a second one, based on the baryon number conservation law and impinging neutron energies about 1.3-2 MeV below the threshold of the corresponding (n,2n) nuclear reaction. Currently, there is no possibility to directly probe the dineutron within the potential well of the residual nucleus. Therefore, we can only rely on the detection of the induced activity of the residual nucleus itself and study the transformations and possible strong interaction of both reaction products, namely the residual nucleus and the dineutron, in time.

**First** expected transformation would be a radioactive decay of the dineutron as a neutron excess nucleus. The only possible decay mode of a bound dineutron, is the $\beta^-$ decay [23, 24]. Otherwise, an additional source of energy is needed for its breakup in order to observe separate neutrons. Then, as a result of the $\beta^-$ decay of the dineutron, we may expect electrons that are leaving the irradiated sample, to be further detected with a corresponding beta-counting technique. For such detection experiment we need to know at least a preliminary estimate of a beta-spectrum end-point energy. This value is in a strong conjunction with another very important nuclear characteristic: the half-life of a bound dineutron, also essential for our study. Both these values are estimated below, based on a very well verified up-to-date approach.

**Second** expected transformation of the residual nucleus-bound dineutron nuclear system may be due to the conversion of the residual nucleus with Z charge into its isobar with Z-1 charge because of the weak interaction between the electron, originating from the $\beta^-$-decay of a bound dineutron, and the residual nucleus. In this study, we show that such a process betwixt the electron and the residual nucleus might indeed take place and its probability $P$ does not equal zero.

**Third** expected transformation refers to the unique nuclear system, that consists of a heavier nucleus ($^{158g}$Tb/$^{158}$Gd/$^{158}$Dy) and a lighter one (the deuteron, as a decay product of a bound dineutron), as a particle-satellite. This nuclear configuration is to some extent similar to the Earth-Moon "double-planet" system. Because of the very small distance (~2 fm) between the deuteron and a heavier nucleus within its potential well, we may expect for an occurrence of the strong interaction between these nuclei, resulting in fusion of $^{158}$Tb/$^{158}$Gd/$^{158}$Dy with the deuteron, and leading to the additional accumulation of $^{160}$Tb, $^{160}$Dy and/or $^{160}$Ho nuclei in a sample. This expectation is based on a similarity of such heavy nucleus-deuteron system to an equivalent two nuclei configuration in a nuclear reaction channel with the impinging deuteron being of certain energy above the reaction threshold, immediately behind the Coulomb barrier and near the surface of this heavy nucleus. The only difference in our experiment is that the deuteron was formed at the opposite side of the Coulomb barrier with a kinetic energy lower than what is needed to reach this location in a close proximity to a heavier nucleus. The first signs of a possible nuclear fusion between such nuclei were noticed in [24, 25]. In this paper, we would also like to stress that the change of $^{160}$Tb/$^{160}$Dy radioactivity in time, observed by means of detection of the 879.3 keV gamma line of $^{160}$Dy, formed directly or as a daughter nucleus of $^{160}$Tb decay, is not smooth. Moreover, this dependence has a maximum at roughly about 440±280 d [25] since the neutron irradiation of the $^{159}$Tb sample was completed on December 6, 2013 at the IRSN facility AMANDE, Cadarache [19]. The markers of nuclear fusion were the following: enhanced activities of the $^{160}$Tb/$^{160}$Dy isotopes and a greater estimated half-life in comparison with the 72.3 d half-life reference value for $^{160}$Tb.

In this study, we attempt to describe the experimental data available and to explain the presence of a maximum in the $^{160}$Tb/$^{160}$Dy radioactivity at about 495 d since the end date of the $^{159}$Tb-sample neutron irradiation.

## 2. Experimental data

All experimental measurements in this research are considered for the same Tb sample, used to determine the $^{159}$Tb(n, γ)$^{160}$Tb nuclear reaction cross section for a 6.85 MeV neutron energy [26]. Information about six measurements of interest is summarized in Table 1. The gamma-line of 879.38 keV ($k_{\gamma 2}$=0.301) of $^{160}$Dy due to the $^{160}$Tb $\beta^-$-decay was used in our research because of no background interference. Two spectrometers with HPGe detectors were utilized for this study, namely, GX4019 at Kyiv Institute for Nuclear Research of National Academy of Sciences of Ukraine (KINR); and GC2020 at the Department of Nuclear Physics, Taras Shevchenko National

University of Kyiv, Ukraine (NUK). Additional data in Table 1 is as follows: $T_{cool.}$ - cooling time from the date of neutron irradiation completion till the end of corresponding counting; $T_{count.}$ - live counting time; $S_p$ - 879.38 keV gamma-line peak area detected in the instrumental gamma-spectrum; $\Delta S_p$ – gamma-line peak area uncertainty. The first instrumental spectrum for this study was acquired ~12 days after the end of the Tb sample neutron irradiation, the last one – about 2.3 years later, before the detection limit was reached for the NUK CANBERRA HPGe gamma-ray spectrometer to reliably observe the 879.38 keV gamma-line peak. Several background spectra, acquired with different counting times, confirmed no significant peak areas detected within the 875÷885 keV energy region of interest. As stated above, besides the $^{159}$Tb $(n,\gamma)$ nuclear reaction product, our measurements included also the study of the $^{159}$Tb$(n,^2n)^{158g}$Tb nuclear process [19], later evincing any possible transformation of the reaction products [24]. In particular, for our calculations we checked the intensity of the 944.2 keV gamma line of the $^{158g}$Tb nucleus.

From our repeatedly processed spectrometric data in Figs. 1-5, being the basis for Table 1, experimental values were determined for the $^{160}$Tb/$^{160}$Dy intensities according to the algorithm, described in [24], and presented along with calculated ones in Table 2.

Table 1. Results of Tb-sample countings.

| No. of count. | HPGe spectrometer / location | $T_{cool.}$, d | Start date of measurement | $T_{count.}$, live, s[b] | $S_p$, counts | $\Delta S_p$, counts |
|---|---|---|---|---|---|---|
| 1. | GX4019/KINR | 12.375 | 18 Dec 2013 | 23,223.14 | 3,244 | 59 |
| 2. | GX4019/KINR | 434.09 | 13 Feb 2015 | 602,386.59 | 2,107 | 77 |
| 3. | GC2020/NUK | 525.2112 | 15 May 2015 | 448,449.10 | 518 | 30 |
| 4. | GC2020/NUK | 575.0037 | 04 July 2015 | 2,003,882.66 | 1,401 | 68 |
| 5. | GC2020/NUK | 624.00 | 22 Aug 2015 | 1,056,547.79 | 469 | 54 |
| 6. | GC2020/NUK | 864.3324 | 18 Apr 2016 | 235,386.43 | 58 | 22 |

Table 2. Results of intensity calculations.

| No. of count. | $S_p/T_{count.}$, cps | $\Delta S_p/T_{count.}$, cps | Intensity, cps | $\Delta$ Intensity, cps | Intensity_1 calculated, cps | Intensity_2 calculated, cps |
|---|---|---|---|---|---|---|
| 1. | 0.140 | 0.003 | 5.35 | 0.39 | 5.38 | 5.75 |
| 2. | 0.0035 | 0.0001 | 0.13 | 0.01 | 0.12 | 0.46 |
| 3. | 0.0012 | 0.0001 | 0.09 | 0.01 | 0.07 | 0.41 |
| 4. | 0.0007 | 0.00004 | 0.056 | 0.006 | 0.050 | 0.39 |
| 5. | 0.00044 | 0.00005 | 0.036 | 0.005 | 0.041 | 0.38 |
| 6. | 0.00025 | 0.00009 | 0.020 | 0.008 | 0.028 | 0.36 |

[b] Dead time for all measurements did not exceed 0.05%

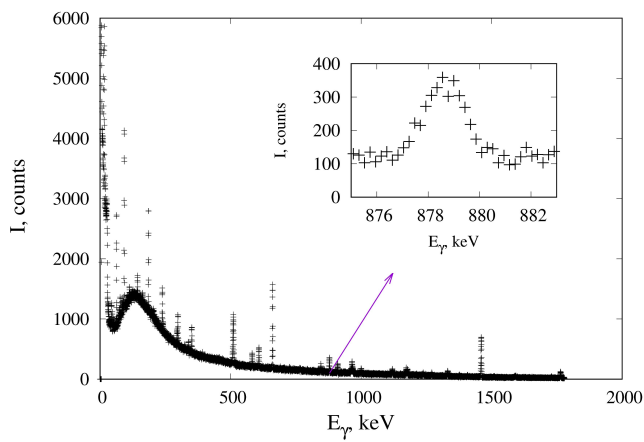

Fig. 1. The instrumental γ-ray spectrum 2 from Table 1.

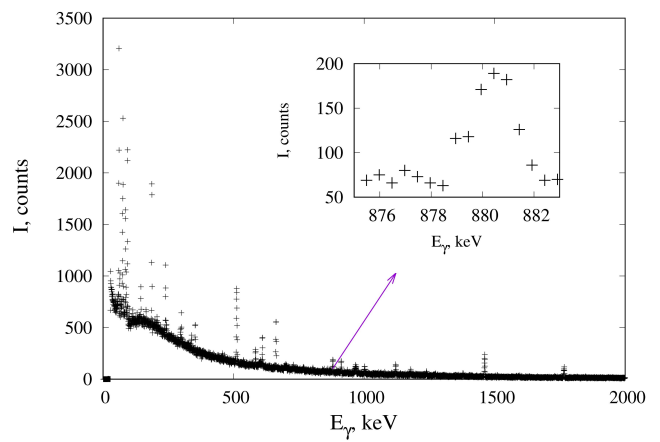

Fig. 2. The instrumental γ-ray spectrum 3 from Table 1.

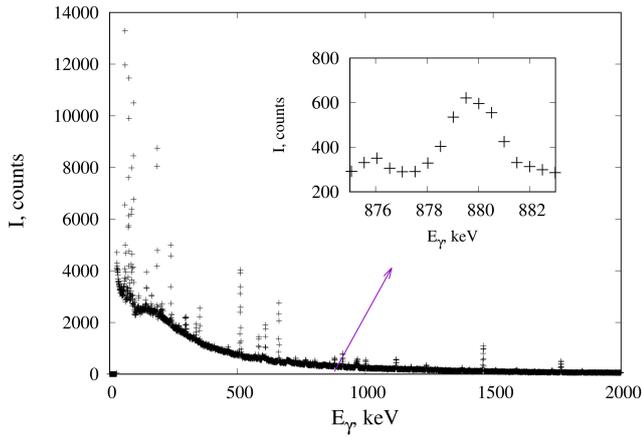

Fig. 3. The instrumental γ-ray spectrum 4 from Table 1.

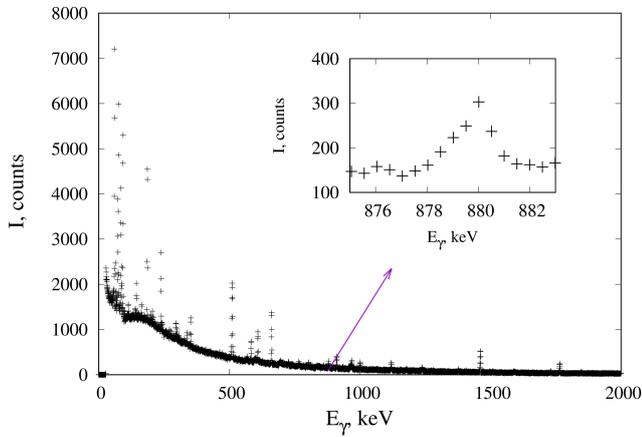

Fig. 4. The instrumental γ-ray spectrum 5 from Table 1.

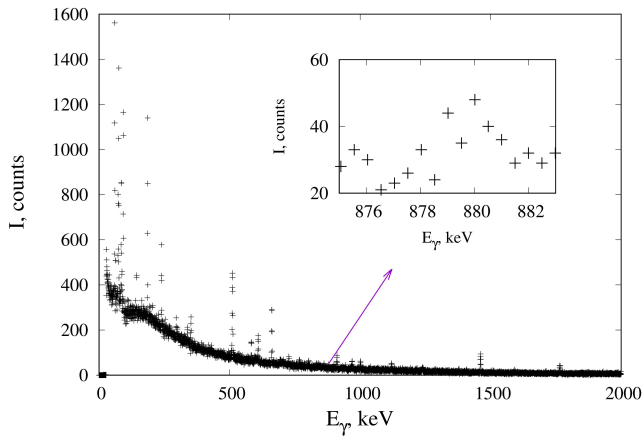

Fig. 5. The instrumental γ-ray spectrum 6 from Table 1.

Data on intensity calculations (columns 4 and 5 of Table 2) were then fitted with the exponential (Fig.6) and the Ln-linear (Fig.7) functions (red) to derive a modified decay constant ($T^m$) for this fusing-decaying system in comparison with a theoretical curve (dashed blue), beginning from the initial point, corresponding to the induced radioactivity of $^{160}$Tb at the end of Tb sample neutron irradiation ($T_{cool}$=0).

At the same time, the first point was omitted from the fitting because of more than 99% contribution due to the decay of $^{160}$Tb, activated in the $^{159}$Tb (n,γ) nuclear reaction [26].

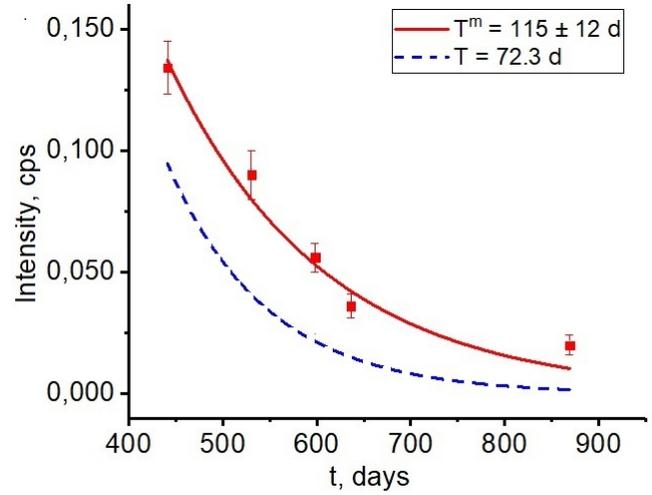

Fig. 6. Experimental intensities fitted with the exponential function.

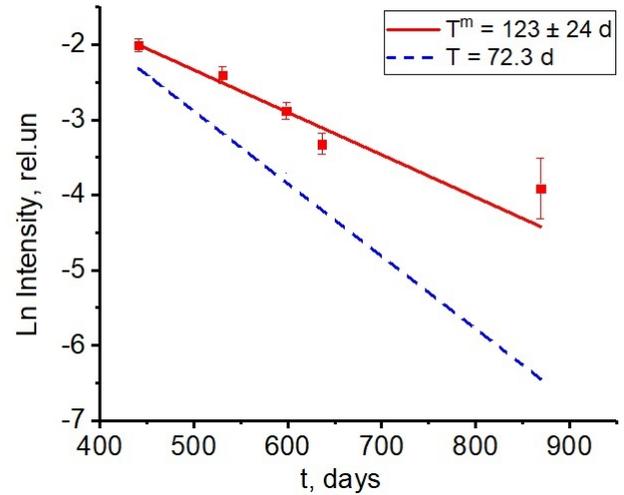

Fig. 7. Ln of experimental intensities fitted with the linear function.

The estimations from the two fittings (115±14 d and 123±24 d) overlap within one sigma, which proves the robustness of the obtained results. Also, even our lesser previous estimate of this decay constant, which is equal to $97^{+16}_{-12}$ d [25, 27], was considered as an outlier [27].

On one hand, we can't help but agree that this is a correct statement of the author [27], yet on another hand, this proves a significant deviation of our two findings of the decay constant from the tabulated value of the half-life [27] and confirms our assumption about the presence of additional radioactive nuclei causing the greater value of the decay constant $T^m$.

As mentioned above, among other nuclear transformations we have to begin considering the radioactive decay of the dineutron as a neutron excess nucleus.

Then for our further calculations, we need to make a reasonable estimate of the dineutron decay constant.

## 3. Half-lives of a bound dineutron

In a very first approximation, we may follow an approach, according to which the dineutron is assumed to be loosely bound but decaying into the deuteron, electron and the electron antineutrino. To estimate its decay constant, one can use the following expression to describe the allowed and superallowed transitions [28]:

$$f_{dn} \cdot t_{dn} = \frac{\tau_{1/2}}{B(F) + \lambda_A^2 \cdot B(GT)}, \quad (1)$$

where: $f_{dn}$ – the phase space factor for the dineutron; $t_{dn}$ - the half-life of the dineutron; $\tau_{1/2}$=6,145 s; $B(F)$ – the Fermi strength; $B(GT)$ – the Gamow-Teller strength; $\lambda_A$=1.27. If we consider the dineutron in a singlet state, decaying into the deuteron in a triplet state, then the Fermi transition is forbidden, i.e. $B(F)$=0. The Gamow-Teller transition is allowed and we may use $B(GT)$=1. Then we need to make an estimate of the phase space and this can be done with the service available by ref. [29]. The result of this estimation gives Log ($f_{dn} \cdot t_{dn}$)=2.104 and this means that ($f_{dn} \cdot t_{dn}$)=127 s. For $t_{dn}$ =1 s we get $f_{dn}$=127. Thus, for the above fixed parameters we apply Eq. (1) and obtain **$t_{dn-1}$=30 s**. This estimation looks interesting from the point of view of theoretical calculations of the expected order of value for the dineutron decay constant and can serve as a lower limit. On another hand, based on the prediction in [4], allowing to compensate at least 66 keV in the binding energy of the dineutron by means of overlapping potential wells of a heavy nucleus and the dineutron, as well as our experimental results and estimates [19, 22, 23], the binding energy of the dineutron does not equal zero. Moreover, in analogy to the isospin formalism, the binding energy of the deuteron and a bound dineutron should be similar or even a bit greater for the dineutron, but cannot exceed 2.5 MeV, *i.e.*

$B_{dn} \lesssim 2.5$ MeV as a reliable upper limit set by BBN [30]. To perform further calculations, we will assume that the binding energy of the dineutron $B_{dn}$ equals 2.45 MeV for an upper estimate of the dineutron half-life. Moreover, this value of the binding energy does perfectly fit our experimentally obtained interval estimate [2.2-2.8] MeV [23].Then in order to make a reasonable assessment of the dineutron half-life, we should assume that for a bound dineutron with $T$=1, $S$=0 and in the state $L$=0 the radial wave function of the dineutron is equivalent to the radial wave function of the deuteron. This assumption can be justified because an expected radius of a bound dineutron in 4.1 fm [23] is more than comparable to the one for the deuteron: 4.3 fm. Of course, the state with $L$=2 for the deuteron is neglected for this case.

Then one can take into account the fact that for the Gamow-Teller transition the sum rule (expression (6.69) in [28]) may be applied and, accordingly, because of $B(GT+)$=0, the maximum $B(GT-)$=6. Then from (1) ($f_{dn} \cdot t_{dn}$) = 635 s. At this stage we also need to know the end-point energy ($E_{max-dn}$) for the $\beta^-$-spectrum of dineutron decay. The very first upper estimate of the binding energy of the dineutron was reported in [31] and equals now 3.01 MeV. Actually, this upper estimate is the sum of the binding energy of the dineutron and the end-point energy of the $\beta^-$-spectrum of electrons due to decay of dineutrons. Then $E_{max-dn}$= 0.56 MeV and we may get $f_{dn}$ from the semi-empirical expression (2) for the phase space factor of the dineutron [23, 32] with the atomic number of the product nuclide $A_d$ = 2:

$$\text{Log } f_{dn} = 4.0 \cdot \text{Log } E_{max-dn} + 0.78 + \\ + 0.02 \cdot A_d - 0.005 \cdot (A_d - 1) \cdot \text{Log } E_{max-dn}. \quad (2)$$

Before doing this calculation, it would be worthwhile to compute Log ($f \; t$) values with application of Eq. (2) and compare them with those from [29] for neighboring neutron and tritium decays. Then if we set for neutron decay end-point energy 0.78232 MeV, $A_d$ = 1 and the 611 s half-life, we get 3.1596 and 3.015, correspondingly. For tritium end-point energy 0.01859 MeV, $A_d$ = 3 and half-life 12.323 y we obtain 2.524 and 3.052, accordingly. As we can compare these estimates, they differ significantly. Therefore, we decided to slightly modify two multiplication factors in Eq. (2) to have Log ($f \cdot t$) values (3.01498 for the neutron and 3.0522 for tritium) now in excellent agreement from the expression below:

$$\text{Log } f_{dn} = 4.0 \cdot \text{Log } E_{max-dn} + 0.6354 + \\ + 0.02 \cdot A_d - 0.1993 \cdot (A_d - 1) \cdot \text{Log } E_{max-dn}. \quad (3)$$

Then from Eq. (3) we get $f_{dn}$ = 0.5228, Log ($f_{dn} \cdot t_{dn-2}$) = 2.803 and finally **$t_{dn-2}$ = 1,215 s**. This transition is the superallowed one. It is worth noting that such an estimate seems reasonable, but one question remains unanswered – what mechanism keeps these fusing-decaying systems running for years? If the Gamow-Teller transition occurs, then the deuteron in a triplet state appears, that might react with $^{158}$Tb or $^{158}$Gd or $^{158}$Dy within a limited time after its formation or won't react at all. Our experimental observations support another idea [24], according to which the deuteron could also be formed in a singlet state, captured at and still occupying one of Migdal's levels in the potential well of $^{158}$Tb/$^{158}$Gd/$^{158}$Dy nuclei. Such system may exist much longer and for this particular case the deuteron has $T$=0, $S$=0, $L$=0 and only the Fermi transition is therefore allowed for the dineutron decay. Then $B(F)$=2 [28] and following the same steps as above, we get Log($f_{dn} \cdot t_{dn-3}$)

=3.487 with $t_{dn-3}$ = **5,877 s**. This transition is rather the allowed one.

Now we have three estimates of the half-life of the dineutron, and the right selection for our subsequent calculations would be the last one as the greatest, compared to other two as it allows both for dineutron and deuteron existence in a singlet state.

## 4. Mathematical model for fusing-decaying nuclear systems

Our mathematical model that describes fusing-decaying systems composed of dineutron/deuteron and $^{158}$Tb/$^{158}$Gd/$^{158}$Dy nuclei consists of three differential eq. systems and is presented below.

### 4.1 System 1 of differential equations

System 1 describes the decay of bound dineutrons (differential eq. 1 below); interaction of electrons, originating from dineutron decays, with $^{158}$Tb nuclei, decay of $^{158}$Tb and fusion of $^{158}$Tb nuclei with deuterons as another dineutron decay product (differential eq. 2 below); accumulation of $^{160}$Dy nuclei because of fusion between $^{158}$Tb and the deuteron (differential eq. 3 below):

$$\begin{cases} \dfrac{dN_{dn}(t)}{dt} = -\lambda_{dn} \cdot N_{dn}(t) \\ \dfrac{dN_{Tb8}(t)}{dt} = -\lambda_{dn} \cdot P \cdot N_{dn}(t) - (\lambda_{Tb8} + F_1) \cdot N_{Tb8}(t) \\ \dfrac{dN^*_{Dy6}(t)}{dt} = F_1 \cdot N_{Tb8}(t) \end{cases}$$

where: $N_{dn}(t)$ – number of dineutron nuclei vs. time $t$; $N_{Tb8}(t)$ – number of $^{158}$Tb nuclei vs. time $t$; $N^*_{Dy6}(t)$ – number of $^{160}$Dy nuclei vs. time $t$; $P$ – probability of $^{158}$Tb transformation into $^{158}$Gd due to the weak interaction with an electron originating from the dineutron decay; $\lambda_{dn}$ – dineutron decay constant; $\lambda_{Tb8}$ – $^{158g}$Tb decay constant; $F_1$ – a fusion constant, describing fusion between $^{158g}$Tb and the deuteron, leading to $^{160}$Dy formation.

System 1 has the corresponding solutions below under the following initial conditions at the moment of irradiation end:

$N_{dn}(0) = N^0_{dn} = N_{Tb8}(0) = N^0_{Tb8} = 2.7 \cdot 10^8/(1-P)$ [5]; $N_{Dy6}(0) = 0$

$N_{dn}(t) = N^0_{dn} \cdot \exp[-\lambda_{dn} \cdot t]$; (solution 1-1)

$N_{Tb8}(t) = \omega \cdot (\exp[-\lambda_{dn} \cdot t] - \exp[-(\lambda_{Tb8} + F_1) \cdot t]) + N^0_{Tb8} \cdot \exp[-(\lambda_{Tb8} + F_1) \cdot t]$; (solution 1-2)

$N^*_{Dy6}(t) = F_1 \cdot (\dfrac{\omega}{\lambda_{dn}} \cdot (1 - \exp[-\lambda_{dn} \cdot t]) + \dfrac{(N^0_{Tb8} - \omega)}{(\lambda_{Tb8} + F_1)} \cdot (1 - \exp[-(\lambda_{Tb8} + F_1) \cdot t]))$, (solution 1-3),

where:

$$\omega = \dfrac{\lambda_{dn} \cdot P \cdot N^0_{dn}}{\lambda_{dn} - (\lambda_{Tb8} + F_1)}.$$

### 4.2 System 2 of differential equations

System 2 describes an increase of the $^{158}$Gd nuclei amount due to absorbed electrons, originating from dineutron decays, by $^{158g}$Tb nuclei and EC/$\beta^+$ decay of $^{158g}$Tb into $^{158}$Gd, as well as diminution of $^{158}$Gd nuclei amount because of fusion with deuterons (differential eq. 1 below); accumulation of $^{160}$Tb nuclei amount as a result of $^{158}$Gd fusion with deuterons and decay of $^{160}$Tb nuclei (differential eq. 2 below); accumulation of $^{160}$Dy nuclei because of a $\beta^-$ decay of $^{160}$Tb nuclei (differential eq. 3 below):

$$\begin{cases} \dfrac{dN_{Gd8}(t)}{dt} = \lambda_{dn} \cdot P \cdot N_{dn}(t) + k_1 \cdot \lambda_{Tb8} \cdot N_{Tb8}(t) - F_2 \cdot N_{Gd8}(t) \\ \dfrac{dN_{Tb6}(t)}{dt} = F_2 \cdot N_{Gd8}(t) - \lambda_{Tb6} \cdot N_{Tb6}(t) \\ \dfrac{dN^{**}_{Dy6}(t)}{dt} = \lambda_{Tb6} \cdot N_{Tb6}(t) \end{cases}$$

where: $N_{Gd8}(t)$ – number of $^{158}$Gd nuclei vs. time $t$; $N_{Tb6}(t)$ – number of $^{160}$Tb nuclei vs. time $t$; $N^{**}_{Dy6}(t)$ – number of $^{160}$Dy nuclei vs. time $t$; $k_1$ – branching ratio of non-affected $^{158}$Tb nuclei disintegrating into $^{158}$Gd according to the $^{158}$Tb decay scheme through EC or $\beta^+$-decay: $k_1 = 0.834$; $F_2$ – a fusion constant, describing the fusion between $^{158}$Gd and the deuteron, leading to the $^{160}$Tb nuclei formation; $\lambda_{Tb6}$ – $^{160}$Tb decay constant.

System 2 has the corresponding solutions below under the following initial conditions:

$N_{Gd8}(0) = N^0_{Gd8} = N_{Dy6}(0) = 0$

$N_{Gd8}(t) = \theta \cdot (\exp[-\lambda_{dn} \cdot t] - \exp[-F_2 \cdot t]) + \xi \cdot (\exp[-(\lambda_{Tb8} + F_1) \cdot t] - \exp[-F_2 \cdot t])$; (solution 2-1)

$$N_{Tb6}(t) = F_2 \cdot \theta \cdot \left( \frac{\exp[-\lambda_{dn} \cdot t] - \exp[-\lambda_{Tb6} \cdot t]}{\lambda_{Tb6} - \lambda_{dn}} - \frac{\exp[-F_2 \cdot t] - \exp[-\lambda_{Tb6} \cdot t]}{\lambda_{Tb6} - F_2} \right) +$$
$$+ F_2 \cdot \xi \cdot \left( \frac{\exp[-(\lambda_{Tb8} + F_1) \cdot t] - \exp[-\lambda_{Tb6} \cdot t]}{\lambda_{Tb6} - (\lambda_{Tb8} + F_1)} - \frac{\exp[-F_2 \cdot t] - \exp[-\lambda_{Tb6} \cdot t]}{\lambda_{Tb6} - F_2} \right);$$

(solution 2-2)

$$N_{Dy6}^{**}(t) = F_2 \cdot \theta \cdot \left( \frac{\lambda_{Tb6} \cdot (1 - \exp[-\lambda_{dn} \cdot t]) - \lambda_{dn} \cdot (1 - \exp[-\lambda_{Tb6} \cdot t])}{(\lambda_{Tb6} - \lambda_{dn}) \cdot \lambda_{dn}} + \frac{F_2 \cdot (1 - \exp[-\lambda_{Tb6} \cdot t]) - \lambda_{Tb6} \cdot (1 - \exp[-F_2 \cdot t])}{(\lambda_{Tb6} - F_2) \cdot F_2} \right) +$$
$$+ F_2 \cdot \xi \cdot \left( \frac{\lambda_{Tb6} \cdot (1 - \exp[-(\lambda_{Tb8} + F_1) \cdot t]) - (\lambda_{Tb8} + F_1) \cdot (1 - \exp[-\lambda_{Tb6} \cdot t])}{(\lambda_{Tb6} - (\lambda_{Tb8} + F_1)) \cdot (\lambda_{Tb8} + F_1)} - \frac{\lambda_{Tb6} \cdot (1 - \exp[-F_2 \cdot t]) - F_2 \cdot (1 - \exp[-\lambda_{Tb6} \cdot t])}{(\lambda_{Tb6} - F_2) \cdot F_2} \right),$$

(solution 2-3),

where:

$$\theta = \frac{\lambda_{dn} \cdot P \cdot N_{dn}^0}{F_2 - \lambda_{dn}} \cdot \left( 1 + \frac{k_1 \cdot \lambda_{Tb8}}{\lambda_{dn} - (\lambda_{Tb8} + F_1)} \right),$$

$$\xi = \frac{k_1 \cdot \lambda_{Tb8}}{F_2 - (\lambda_{Tb8} + F_1)} \cdot \left( N_{Tb8}^0 - \frac{\lambda_{dn} \cdot P \cdot N_{dn}^0}{\lambda_{dn} - (\lambda_{Tb8} + F_1)} \right).$$

### 4.3 System 3 of differential equations

System 3 describes an increase of the $^{158}$Dy nuclei amount due to $^{158g}$Tb nuclei $\beta^-$-decay and also its decrease due to fusion with deuterons (differential eq. 1 below); an accumulation of $^{160}$Ho nuclei due to the fusion of $^{158}$Dy nuclei with deuterons and the decay of $^{160}$Ho nuclei (differential eq. 2 below); the accumulation of $^{160}$Dy nuclei because of $^{160}$Ho EC/$\beta^+$-decay (differential eq. 3 below):

$$\begin{cases} \dfrac{dN_{Dy8}(t)}{dt} = k_2 \cdot \lambda_{Tb8} \cdot N_{Tb8}(t) - F_3 \cdot N_{Dy8}(t) \\ \dfrac{dN_{Ho6}(t)}{dt} = F_3 \cdot N_{Dy8}(t) - \lambda_{Ho6} \cdot N_{Ho6}(t) \\ \dfrac{dN_{Dy6}^{***}(t)}{dt} = \lambda_{Ho6} \cdot N_{Ho6}(t) \end{cases},$$

where: $N_{Dy8}(t)$ - number of $^{158}$Dy nuclei vs. time $t$; $N_{Ho6}(t)$ - number of $^{160}$Ho nuclei vs. time $t$; $N_{Dy6}^{***}(t)$ - number of $^{160}$Dy nuclei vs. time $t$; $k_2$ - branching ratio of non-affected $^{158}$Tb nuclei disintegrating into $^{158}$Dy according to the $^{158}$Tb decay scheme through $\beta^-$ decay: $k_2 = 0.166$; $F_3$ - a fusion constant, describing the fusion between $^{158}$Dy and the deuteron and leading to $^{160}$Ho formation; $\lambda_{Ho6}$ – $^{160m}$Ho decay constant.

System 3 has the corresponding solutions below under the following initial conditions:

$$N_{Dy8}(0) = N_{Dy8}^0 = 0 \ ; \ N_{Ho6}(0) = 0 \ ; \ N_{Dy6}(0) = 0$$

$$N_{Dy8}(t) = \varpi \cdot k_2 \cdot (\exp[-(\lambda_{Tb8} + F_1) \cdot t] - \exp[-F_3 \cdot t]) + \chi \cdot (\exp[-\lambda_{dn} \cdot t] - \exp[-F_3 \cdot t]); \quad \text{(solution 3-1)}$$

$$N_{Ho6}(t) = F_3 \cdot k_2 \cdot \varpi \cdot \left( \frac{\exp[-(\lambda_{Tb8} + F_1) \cdot t] - \exp[-\lambda_{Ho6} \cdot t]}{\lambda_{Ho6} - (\lambda_{Tb8} + F_1)} + \frac{\exp[-F_3 \cdot t] - \exp[-\lambda_{Ho6} \cdot t]}{F_3 - \lambda_{Ho6}} \right) +$$
$$+ F_3 \cdot \chi \cdot \left( \frac{\exp[-F_3 \cdot t] - \exp[-\lambda_{Ho6} \cdot t]}{F_3 - \lambda_{Ho6}} + \frac{\exp[-\lambda_{dn} \cdot t] - \exp[-\lambda_{Ho6} \cdot t]}{\lambda_{Ho6} - \lambda_{dn}} \right);$$

(solution 3-2)

$$N_{Dy6}^{***}(t) = k_2 \cdot \varpi \cdot \left( \frac{F_3}{\lambda_{Ho6} - (\lambda_{Tb8} + F_1)} \cdot \left( \frac{\lambda_{Ho6}}{\lambda_{Tb8} + F_1} \cdot (1 - \exp[-(\lambda_{Tb8} + F_1) \cdot t]) - (1 - \exp[-\lambda_{Ho6} \cdot t]) \right) + \right.$$
$$+ \frac{1}{F_3 - \lambda_{Ho6}} \cdot (\lambda_{Ho6} \cdot (1 - \exp[-F_3 \cdot t]) - F_3 \cdot (1 - \exp[-\lambda_{Ho6} \cdot t]))) + \chi \cdot \left( \frac{1}{F_3 - \lambda_{Ho6}} \cdot (\lambda_{Ho6} \cdot (1 - \exp[-F_3 \cdot t]) - F_3 \cdot (1 - \exp[-\lambda_{Ho6} \cdot t])) + \right.$$
$$+ \left. \frac{F_3}{\lambda_{Ho6} - \lambda_{dn}} \cdot \left( \frac{\lambda_{Ho6}}{\lambda_{dn}} \cdot (1 - \exp[-\lambda_{dn} \cdot t]) - (1 - \exp[-\lambda_{Ho6} \cdot t]) \right) \right),$$

(solution 3-3),

where:

$$\chi = \frac{k_2 \cdot \lambda_{Tb8}}{F_3 - \lambda_{dn}} \cdot \frac{\lambda_{dn} \cdot P \cdot N_{dn}^0}{\lambda_{dn} - (\lambda_{Tb8} + F_1)},$$

$$\varpi = \frac{\lambda_{Tb8}}{F_3 - (\lambda_{Tb8} + F_1)} \cdot \left( N^0_{Tb8} - \frac{\lambda_{dn} \cdot P \cdot N^0_{dn}}{\lambda_{dn} - (\lambda_{Tb8} + F_1)} \right).$$

### 4.4 Fusion parameters $F_1$ - $F_3$ and probability P

We now consider how to determine the fusion parameters $F_1$ - $F_3$ and the probability $P$, starting with $F_1$. This fusion parameter can be determined to meet the following criteria: due to $^{158g}$Tb decay into $^{158}$Gd, the 944.2 keV gamma-line peak count rate must be equal $1.6 \cdot 10^{-4}$ 1/s [19] for the very last counting No.6, Table 1, as it was experimentally observed in the instrumental gamma-ray spectrum. This count rate, divided by 944.2 keV gamma rays efficiency and the corresponding 944.2 keV gamma-line yield, gives us the value of the $^{158g}$Tb 944 keV intensity, which is equal to the $^{158g}$Tb induced activity in our sample. Then, taking into account the very well-known relation between the radioactivity value and the number of corresponding nuclei, we immediately get the number of $^{158g}$Tb nuclei. Now the left part of the solution 1-2 is determined. In the right part of the same expression, the decay constants, the time $t$ and other parameters are defined and described above and below. By varying $F_1$ we can easily reach an equality of the right and the left sides of this expression. This step does identify $F_1 = 1.4 \cdot 10^{-9}$ 1/s.

To make an estimate of $F_2$, we may use, as a very first approximation, eq. No. 2 of System 2. It is well-known that this eq. is of the same mathematical form as the one that describes the amount of nuclei in the ensemble, consisting of the parent and daughter nuclei in chain. In our particular case, the "parent" part is not the decay, but the fusion of $^{158}$Gd nuclei with deuterons, resulting in $^{160}$Tb nuclei accumulation, and the "daughter" part represents the decay of $^{160}$Tb nuclei. The form of this differential eq. is similar to well-known parent-daughter nuclear decay system, which has a solution with a maximum (in our case for $^{160}$Tb nuclei) vs. time, and Eq. (4) below gives a time moment $T_{max}$, for which an accumulation of daughter nuclei $^{160}$Tb reaches a maximum value, then decreases and follows the "decay" of parent nuclei:

$$T_{max} = \text{Ln}(\lambda_{Tb6}/F_2)/(\lambda_{Tb6} - F_2). \quad (4)$$

For $T_{max}$ = 440 d [25] we get $F_2 = 1.74 \cdot 10^{-9}$ 1/s. Later, based on our data from Table 1, the value of $T_{max}$ was precised and set on 495±8 d fixing a slightly modified value for $F_2 = 1.89 \cdot 10^{-9}$ 1/s.

For the determination of the parameter $F_3$ we applied a similar approach like the one for the parameter $F_2$ and found out that for a reasonable range of $F_3$ parameter ($[1 \cdot 10^{-9} \div 1 \cdot 10^{-13}]$ 1/s) the maximum in $^{160}$Ho activity was not identified. This feature can be explained by the short half-life of $^{160m}$Ho (5.02 h) and due to the fact that the accumulation of $^{160}$Ho is based on the amount of $^{158}$Dy as product nuclei due to the $^{158g}$Tb $\beta^-$-disintegration. Because of the low amount of $^{158}$Dy in our sample, there will be no significant influence at the 879.38 keV gamma-line intensity due to this fusion-decay channel. Moreover, this system of nuclei will be in a secular equilibrium, i.e. per one formation of $^{158}$Dy we can expect a minimal number of $^{160}$Ho decays with 879.38 keV gamma-rays irradiation. Then based on our expectations, we accepted $F_3 = 9 \cdot 10^{-10}$ 1/s. Even if all these fusion parameters are described sequentially in line with their determination, they were calculated simultaneously all together with another parameter $P$.

Now let's move to the probability $P$. This parameter can be derived from the second eq. of the System 2 when the right part of this eq. equals zero in the extremum (maximum) point. Then we get the following expression for $P = f(F_1, F_2)$

$$P = \eta \cdot \left( \frac{\lambda_{dn} \cdot \eta}{(\lambda_{dn} - \lambda_{Tb8} - F_1)} + \frac{\lambda_{dn}}{(F_2 - \lambda_{dn})} \cdot \left(1 + \frac{k_2 \cdot \lambda_{Tb8}}{\lambda_{dn} - \lambda_{Tb8} - F_1}\right) \cdot \left(\frac{F_2}{\lambda_{Tb6} - F_2} \cdot \exp[-F_2 \cdot t] - \left(\frac{1}{\lambda_{Tb6} - F_2} - \frac{1}{\lambda_{Tb6} - \lambda_{dn}}\right) \cdot \lambda_{Tb6} \cdot \exp[-\lambda_{Tb6} \cdot t])\right)^{-1}, \quad (5)$$

where:

$$\eta = \frac{k_2 \cdot \lambda_{Tb8}}{F_2 - \lambda_{Tb8} - F_1} \cdot \left( \frac{\lambda_{Tb8} + F_1}{\lambda_{Tb6} - \lambda_{Tb8} - F_1} \right) \cdot \exp[-(\lambda_{Tb8} + F_1) \cdot t] - \left( \frac{F_2}{\lambda_{Tb6} - F_2} \right) \cdot \exp[-F_2 \cdot t] + \left( \frac{1}{\lambda_{Tb6} - F_2} - \frac{1}{\lambda_{Tb6} - \lambda_{Tb8} - F_1} \right) \cdot \lambda_{Tb6} \cdot \exp[-\lambda_{Tb6} \cdot t])$$

Substituting values of $F_1$ and $F_2$ as well as other known parameters into the Eq. (5) above, we get the following estimate: $P = 0.1017$.

### 4.5 Peak intensities determination

Before doing this set of calculations, we added to the $N_{Tb6}(t)$ expression a member, dealing with a certain amount of nuclei of $^{160}$Tb due to the (n, γ) reaction on $^{159}$Tb [24, 26]. Now, having available dependences for $N^*_{Dy6}(t)$ from the System 1, $N^{**}_{Dy6}(t)$ from the System 2 and $N^{***}_{Dy6}(t)$ from the System 3, we applied the following eq. to calculate the intensity of the 879.38 keV gamma-line:

$$I = \varepsilon_d \cdot \{[N^*_{Dy6}(T_{cool}+T_{count}) - N^*_{Dy6}(T_{cool})] \cdot \frac{k_{\gamma 1}}{T_{count}} + [N^{**}_{Dy6}(T_{cool}+T_{count}) - N^{**}_{Dy6}(T_{cool})] \cdot \frac{k_{\gamma 2}}{T_{count}} + \\ + [N^{***}_{Dy6}(T_{cool}+T_{count}) - N^{***}_{Dy6}(T_{cool})] \cdot \frac{k_{\gamma 3}}{T_{count}}\} \quad (6)$$

where: $\varepsilon_d$ – detection efficiency of 879.38 keV gamma-line $k_{\gamma 1}$ – the transition intensity of the 879.38 keV gamma-line due to direct fusion between $^{158}$Tb nuclei with the deuterons leading to the direct formation of $^{160}$Dy in one of its excited states; $k_{\gamma 2}$ is defined above; $k_{\gamma 3} = 0,21168$ - quantum yield of 879.38 keV line of $^{160}$Dy due to $^{160m}$Ho decay. Actually, $k_{\gamma 1}$ (100%) from the TOICD database is not applicable because of an essential discrepancy between intensity $I$ calculated (Intensity_2) with experimental data (Intensity and Δ Intensity), see corresponding values in Table 2: column 7 and columns 4 and 5. Then, by fitting calculated $I$ values to experimental ones, we found $k_{\gamma 1} = 0.03$. The results of this finding are presented again in Table 2, column 6 (Intensity_1). Now we observe a very good agreement between experimental and calculated data with all parameters of our mathematical model fixed: $F_1$, $F_2$, $F_3$, $P$ and $k_{\gamma 1}$.

### 4.6 Other half-life and miscellaneous calculations

With the development of our mathematical model, we are ready now to calculate several values necessary to deeply understand this very unusual nuclear physical process. To do so, we can take as a reference the measurement No. 1 from Table 1 to verify our algorithm by calculating the half-life of $^{160}$Tb. With the application of Eq. (7) below we got the following result: 72.5 d to be in excellent agreement with the reference value 72.3(2) d. Now our model is verified and we can perform further calculations.

Firstly, from this measurement we can obtain a value of a modified half-life for $^{160}$Tb isotope in days from the measurement No.4 in Table 1 as a middle point with the greatest acquisition time and the Eq. (7) below:

$$T^m_{1/2} = \frac{\ln(2) \cdot T_{count}}{\ln\left(\frac{N_{Tb6}(T_{cool} \cdot 24 \cdot 3600)}{N_{Tb6}(T_{cool} \cdot 24 \cdot 3600 + T_{count})}\right)}. \quad (7)$$

Substituting the corresponding values from Table 1 and the System 2, we get the following modified half-life for $^{160}$Tb: 126.8 d as it was expected in [24]. This value is in a very good agreement with experimental data from Figs. 6 and 7.

Secondly, using the same Eq. (7) we analogously obtain a modified half-life for the $^{158g}$Tb isotope: 14.4 y. This result is more than one order of magnitude less than the 180 y decay constant for this isotope from the nuclear data base [33].

Thirdly, again from the same Eq. (7) we can calculate the "breakup half-life" for $^{158}$Gd, which is expected to be negative because of accumulation, but not a disintegration of this stable isotope of gadolinium due to EC/$\beta^+$ decay of $^{158g}$Tb. What we get is surprisingly opposite: the "breakup half-life" is positive and equals 21.6 y.

Such transformation of the half-life for the $^{158g}$Tb isotope and even the introduction of the "breakup half-life" for the stable $^{158}$Gd isotope are applicable only for this particular nuclear configuration, which is characterized by the presence of the deuteron, as a dineutron decay product, in the close proximity to a heavy nucleus within its potential well. Then the fusion between nuclei in such configuration might occur and possibly followed by the transformation of a heavy nucleus (Z, A) into another one with (Z+1) and (A+2) numbers. Of course, the result of such fusion will greatly depend on the fusion cross section. In case of this fusion cross section being high enough, we can expect and observe an additional mechanism to change the amount of heavy nuclei, irrespective of whether they are stable or radioactive. This mechanism will be lasting until the major part of deuterons will fuse with heavy nuclei, whereupon the original half-lives will be re-established, if any.

Fourthly, all 9 dependences of nuclei amount vs. time as solutions of Systems 1 through 3 are presented in Figs. 8-16.

At first glance, Fig. 9 may have an erroneously high very first data point, but it is not. Because the probability parameter $P$ does not equal zero, the number of $^{158}$Tb nuclei within the very first day after irradiation of the Tb sample diminished sharply due to dineutron decays.

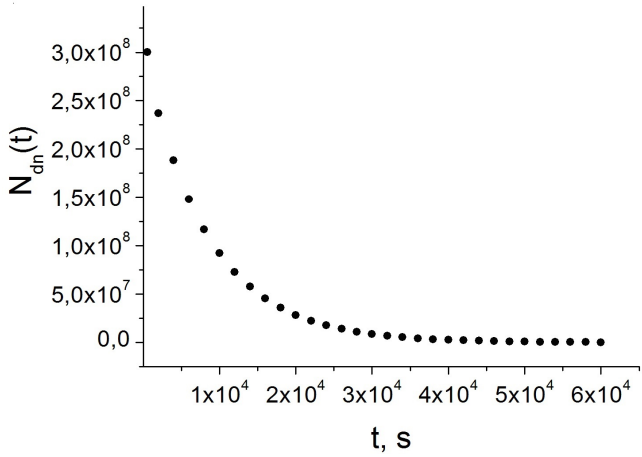

Fig. 8. The dependence of the $N_{dn}(t)$ vs. time.

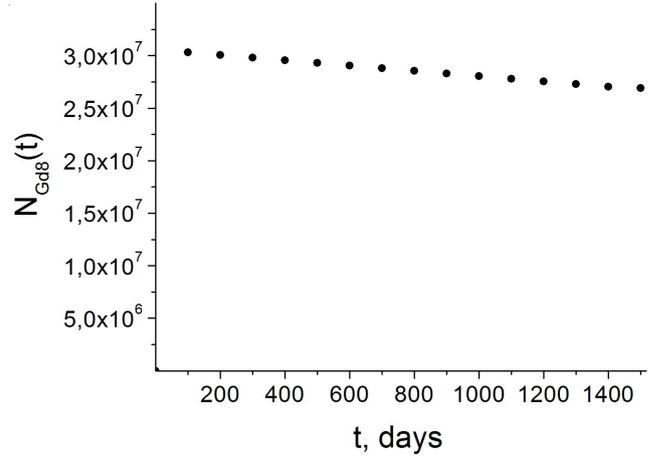

Fig. 11. The dependence of the $N_{Gd8}(t)$ vs. time.

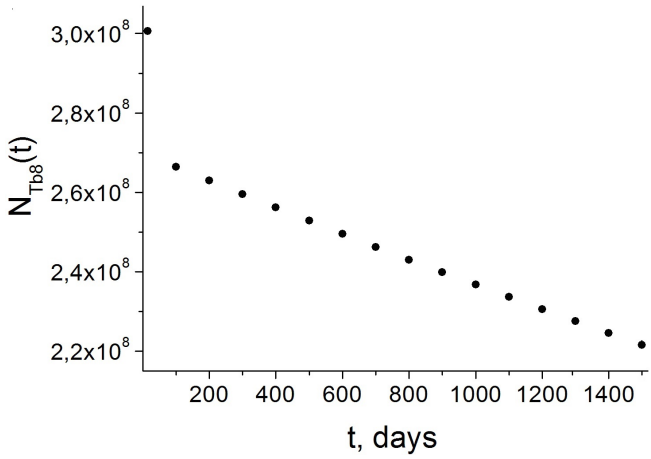

Fig. 9. The dependence of the $N_{Tb8}(t)$ vs. time.

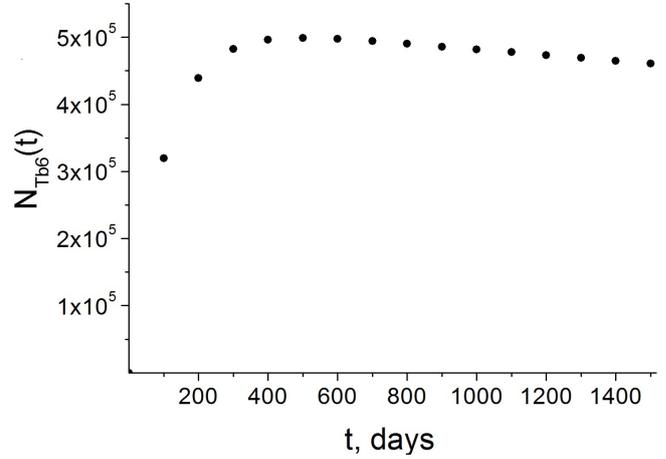

Fig. 12. The dependence of the $N_{Tb6}(t)$ vs. time.

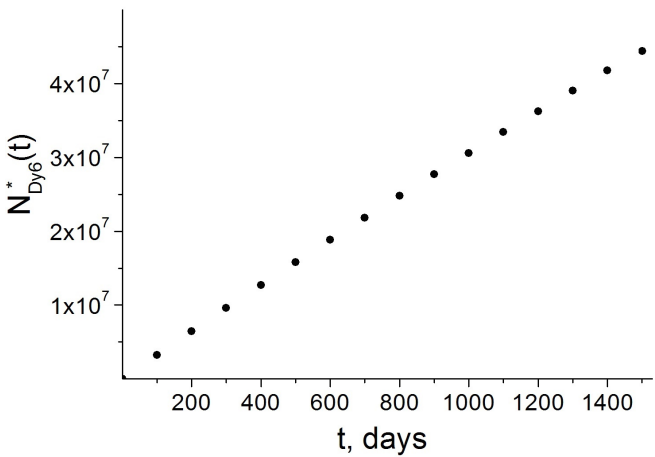

Fig. 10. The dependence of the $N^{*}_{Dy6}(t)$ vs. time.

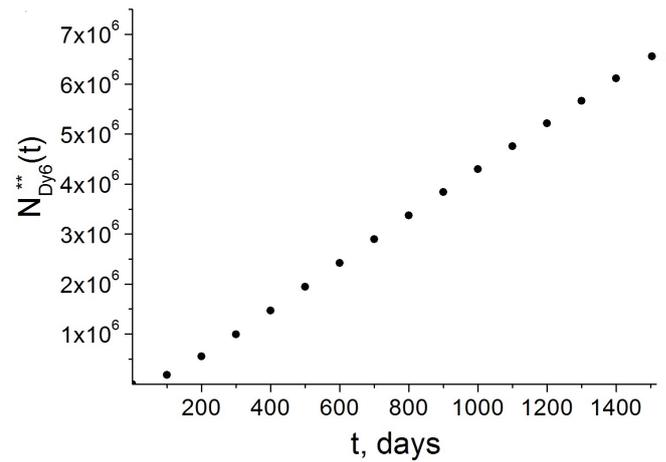

Fig. 13. The dependence of the $N^{**}_{Dy6}(t)$ vs. time.

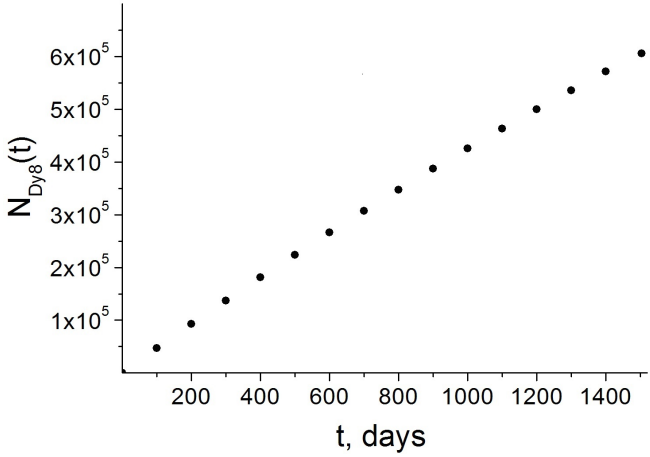

Fig. 14. The dependence of the $N_{Dy8}(t)$ vs. time.

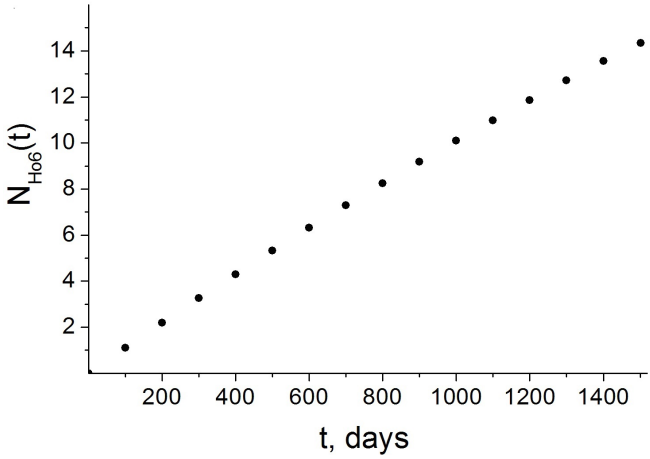

Fig. 15. The dependence of the $N_{Ho6}(t)$ vs. time.

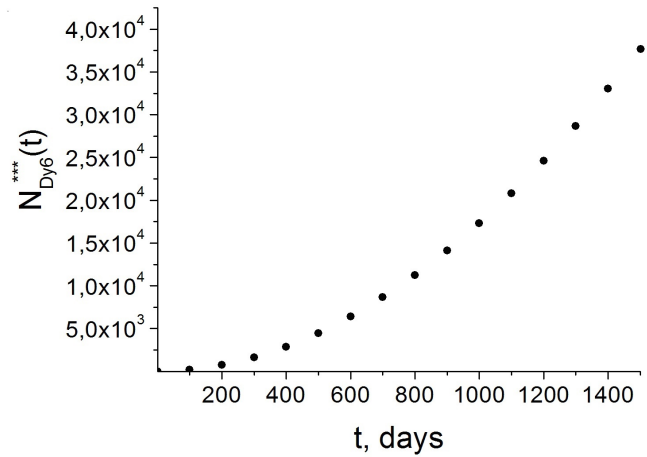

Fig. 16. The dependence of the $N^{***}_{Dy6}(t)$ vs. time.

In Fig. 17 are shown the three separate values from all three systems and total 879.38 keV gamma-peak intensity vs. time with the greatest contribution from System 2 through the accumulation of $^{160}$Tb activity due to fusion between $^{158}$Gd and the deuteron with the identified maximum at about 495 days since Tb sample neutron irradiation completion.

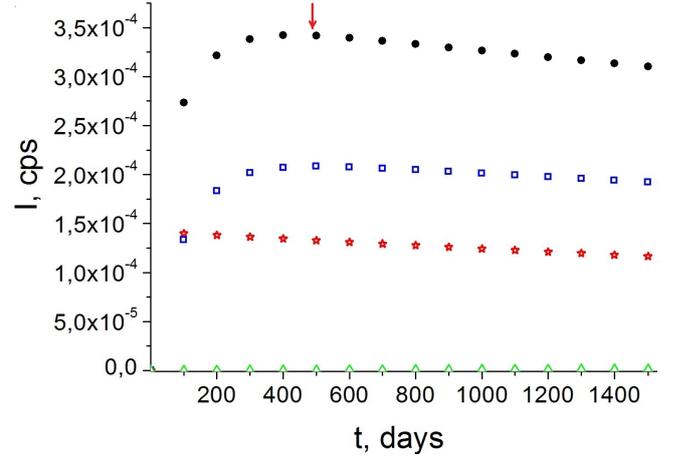

Fig. 17. (color online) The dependence of the intensity of the 879.38 keV gamma-line $I(N^{*}_{Dy6}(t), N^{**}_{Dy6}(t), N^{***}_{Dy6}(t))$ defined in Eq. (6), vs. time. Symbols: black dots - $I(N^{*}_{Dy6}(t)+N^{**}_{Dy6}(t)+N^{***}_{Dy6}(t))$, red stars - $I(N^{*}_{Dy6}(t))$, blue squares - $I(N^{**}_{Dy6}(t))$, green triangles - $I(N^{***}_{Dy6}(t))$. No $(n,\gamma)$ member is taken into account in this fig.

Having these dependences, we can make necessary estimate of another very important parameter to characterize fusion reactions of a lighter nucleus and a heavier ones being in thermal equilibrium under room temperature conditions. This parameter stays for fusion reaction cross section.

## 5. Fusion reaction rates and cross section in thermal equilibrium

We assume that the light nucleus ($d$) and two of the target nuclei $^{158g}$Tb/$^{158}$Gd/$^{158}$Ho, in this case -$^{158}$Tb and $^{158}$Gd, are in thermal equilibrium under room temperature conditions and follow the Maxwell-Boltzmann relative velocity distribution:

$$\Phi(v) = 4\pi \cdot [\mu/(2\pi \cdot k \cdot T_R)]^{3/2} \cdot v^2 \cdot \exp(-\mu v^2/(2k \cdot T_R)),$$

where: $\mu = 0.9 \cdot \mu_{Tb8} + 0.1 \cdot \mu_{Gd8}$ is the reduced mass: $\mu_{Tb8-d} = (m_d \cdot M_{Tb8})/(m_d + M_{Tb8})$; $\mu_{Gd8-d} = (m_d \cdot M_{Gd8})/(m_d + M_{Gd8})$; weights 0.9 and 0.1 represent parts of $^{158g}$Tb and $^{158}$Gd nuclei in these nuclear systems; $k$ is the Boltzmann constant; $T_R$ is a room temperature and $v$ is the relative velocity of a lighter and a heavier nuclei.

Then we can calculate an averaged relative thermal velocity with the following parameters: $T_R = 293.6$ °K; $\mu_{Tb8-d} = 3.30148 \cdot 10^{-27}$ kg; $\mu_{Gd8-d} = 3.30124 \cdot 10^{-27}$ kg and the integral in the second line of Eq. (8) equals:

$$\langle v \rangle = \int_0^\infty \Phi(v) \cdot v \cdot dv = 1.786 \cdot 10^5 \text{ cm/s}.$$

Now a reaction rate *r* for this nuclear fusion process can be expressed as follows:

$$r = N \cdot \frac{N_d}{V} \cdot \int_0^\infty v \cdot \Phi(v) \cdot \sigma_{fus}(v) dv = N \cdot \frac{N_d}{V} \cdot \langle \sigma_{fus} \rangle \cdot \int_0^\infty v \cdot \Phi(v) dv, \qquad (8)$$

where: $N = N_{Tb8} + N_{Gd8}$; $N_d = N$ are numbers of $^{158g}$Tb, $^{158}$Gd and *d* - nuclei vs. time in our sample and these values can be calculated from solutions 1-2 and 2-1 for the corresponding time parameters; *V* – volume of Tb sample [19]; $\sigma_{fus}(v)$ and $\langle \sigma_{fus} \rangle$ – fusion cross section vs. *v* and averaged fusion cross section, accordingly.

Here we need to stress that the fusion between $^{158}$Gd and *d* as well as between $^{158}$Dy and *d* goes via the parent nucleus: $^{160}$Tb and $^{160}$Ho, accordingly, which later must decay in $^{160}$Dy. From Figs. 15 and 17 one can see that the amount of $^{158}$Dy nuclei is negligible in comparison with $^{158g}$Tb and $^{158}$Gd, so we did not take into account a fusion process between $^{158}$Dy and the deuteron as it does not significantly contribute to the reaction rate *r*. Therefore, we decided for our calculations to use only solutions 1-2 and 2-1 from Systems 1 and 2 of differential eqs., describing fusion between $^{158g}$Tb and *d* as well as between $^{158}$Gd and *d*, followed by a direct formation of $^{160}$Dy and $^{160}$Tb, for calculations of necessary amounts of nuclei. Now for all measurements Nos.1-6 from Table 1 we got $^{158}$Tb and $^{158}$Gd numbers of nuclei and subsequently the weighted average estimate of fusion cross section, which is given below:

$\langle \sigma_{fus} \rangle = (1.22 \pm 0.77) \cdot 10^8$ b.

## 6. Discussion

It is well known that nuclear fusion is a process in which at least two nuclei combine to form a heavier nucleus along with a simultaneous release of some amount of energy. For nuclear fusion it is required that the nuclei are forced into close proximity to each other (confinement). Then the attractive nuclear force betwixt nuclei outweighs the electrical repulsion and allows them to fuse. There are several types of confinement in the known fusion mediums: gravitational confinement in stars; magnetic confinements in tokamaks and stellarators; inertial confinement in experiments with laser-induced fusion; lattice confinement in solid bodies.

In our experiment, none of such configurations are present, which brings the need to introduce a new type of confinement: potential well confinement. The occurrence of such a confinement is based on a very specific scenario: contrary to the common approach when charged particles have to be 1-2 fm apart, typical for the strong interaction, in our research the dineutron is formed within the potential well of a heavier nucleus. The dineutron then decays into the deuteron, therefore a charged particle (the deuteron) appears, also being localized within the potential well. The formation of the dineutron in a bound state plays a key role in this nuclear process. Provided that our assumptions and calculations will be experimentally validated by other researches, then a bound dineutron may become the very first nucleus that may decay from its ground state into two different ground states of another nucleus - the deuteron - with two different half-lives: 1,215 s and 5,877 s. Estimated decay constants are great enough to design and perform an experiment with off-line measurements, and hence under favorable experimental conditions.

For our mathematical model we also determined the fusion parameters $F_1$-$F_3$. The most interesting is the fact that the values of all of them are comparable, and $F_2 > F_1 > F_3$. If these parameters are not very different, it means that fusion processes between some heavy nuclei and the deuteron, have common features. Indeed, all the heavy nuclei ($^{158}$Gd, $^{158}$Tb, $^{158}$Dy) are isobars and may behave similarly while fusing. Also, this inequality between the parameters $F_1$- $F_3$ may reflect the fact of more likely fusion process for isobars with lower Z.

Parameter *P* was also calculated, and its value is slightly above 0.1. A non-zero value of this parameter means that the weak interaction between a residual heavy nucleus in the output channel of a nuclear reaction and an electron as a product of $\beta^-$- decay of the dineutron may take place. And it is another type of the weak interaction, not equivalent to the EC mode. Moreover, this fact may be experimentally confirmed, taking into account the decay level scheme and the corresponding gamma-transitions in the $^{158}$Tb nucleus. We also tried to vary the $F_2$ parameter in order to search for its value that corresponds to *P* being as close as possible to zero. Our results are the following: for $P = 2.4738 \cdot 10^{-19}$ we got $F_2 = 3.3939 \cdot 10^{-7}$. This estimate of $F_2$ is about two orders of magnitude greater than $F_1$ and such a huge difference could not be reasonably explained, proving our reliable estimate for *P*.

We also would like to point out a low value of the transition intensity $k_{\gamma 1}$. This estimate proves that the excitation due to the (*d*, *γ*) reaction is insignificant, which seems to be reasonable because of the room temperature conditions and the very low energies of interacting particles. This peculiarity may be promising in order to have a major part of the Q-value of the fusion reaction between $^{158}$Tb and the deuteron (13.3 MeV) in a form of kinetic energy of $^{160}$Dy and not for gamma-irradiation. This expectation is supported by the value of the fusion cross-section which is found to be very high (~$1.2 \cdot 10^8$ b) and ensures the conversion of a significant amount of heavy nuclei due to the fusion process.

And the last interesting result is that the half-life of heavy nuclei that are involved into the corresponding transformations may be reduced significantly and this phenomenon opens up the window of opportunity for potential practical applications, including the transmutation of some fission products.

## 7. Conclusions

In this research, we present experimentally obtained results that allow us to suggest the formation of a bound dineutron in the outgoing channel of the $^{159}$Tb (*n*, $^2$*n*) $^{158g}$Tb nuclear reaction followed by assumed transformations of the reaction products. Current estimations for dineutron half-lives are presented, with 5,877 s being an upper limit.

A reasonable mathematical model that describes the experimental results is provided. A good agreement between experimental and theoretical data adds to the validity of the suggested phenomenon of fusion between heavy nuclei and deuterons under room temperature conditions. However, the suggested model needs to be further expanded, taking into account all the features of the nuclear transformations in the sample.

For a more comprehensive research, further experiments are needed, and the calculations suggest that such experiments can be designed and conducted at existing facilities. The elaboration of a new theoretical approach is also required in order to develop an understanding of the described nuclear systems and their transformations. Namely, the existence of a bound dineutron, that leads to the assumption of the existence of (1) a bound deuteron in a singlet state, (2) decay type, when the same nucleus in the same ground state decays with two different half-lives and (3) nuclear fusion under room temperature conditions, open new opportunities for research of nuclear properties and for their applications.

**Declaration of competing interest**

The authors note that they have no established conflicting financial interests or personal relationships that the work reported in this paper may have appeared to affect.

**Funding**

This research did not receive any specific grant from funding agencies in the public, commercial, or not-for-profit sectors.

**Acknowledgements**

One of the authors (IMK) is sincerely grateful to Prof. A. Volya for very useful discussion of the algorithms to estimate the decay constant of the dineutron. Also the authors are thank Mr. V.Moroziuk for independent checking up differential equation system solutions.